\documentclass[preprintnumbers,article,amsmath,amssymb,floatfix,10pt,prd,superscriptaddress,nofootinbib]{revtex4}

\usepackage{doi}
\usepackage{hyperref}
\hypersetup{
  colorlinks=true,        
  linkcolor=magenta,         
  citecolor=red,         
}

\usepackage{graphicx}
\usepackage{dcolumn}
\usepackage{bm}
\usepackage{color}
\usepackage{enumitem}
\usepackage{amsmath}
\usepackage{amssymb}

\newcommand{\beq}{\begin{equation}}
\newcommand{\eeq}{\end{equation}}
\newcommand{\bea}{\begin{eqnarray}}
\newcommand{\eea}{\end{eqnarray}}

\usepackage{bbm}
\usepackage{amsfonts}
\usepackage{mathrsfs}
\usepackage{latexsym}
\usepackage{epsfig}
\usepackage{epstopdf}
\usepackage{epstopdf}
\usepackage{graphicx}
\usepackage{amssymb}
\usepackage{mathtools}
\usepackage{dcolumn}
\usepackage{bm}
\usepackage{color}
\usepackage{comment}
\usepackage{xcolor}
\usepackage{xcolor}
\usepackage{orcidlink}

\begin{document}
\title{Generalized Exact Fractional Quantum Information Model with Memory Effects}

\author{Abdelmalek Bouzenada\orcidlink{0000-0002-3363-980X}}
\email{abdelmalekbouzenada@gmail.com}
\affiliation{Laboratory of Theoretical and Applied Physics, Echahid Cheikh Larbi Tebessi University, 12001 Tebessa, Algeria}
\affiliation{Research Center of Astrophysics and Cosmology, Khazar University, AZ1096 Baku, 41 Mehseti Street, Azerbaijan}

\author{Allan R. P. Moreira\orcidlink{0000-0002-6535-493X}}
\email{allan.moreira@fisica.ufc.br (Corresp. author)}
\affiliation{Secretaria da Educa\c{c}\~ao do Cear\'a (SEDUC), Coordenadoria Regional de Desenvolvimento da Educa\c{c}\~ao (CREDE 9), Horizonte, Cear\'a 62880-384, Brazil}

\begin{abstract}
In this paper, we analyze quantum information measures in fractional quantum mechanics using the Riemann-Liouville derivative formalism adopted here. In this case, we initially reconsider the conventional definitions of Shannon entropy and Fisher information, subsequently extending them to fractional quantum systems described by nonlocal differential operator frameworks adopted. Within this generalized formulation, fractional expressions of Shannon entropy and Fisher information are constructed and their mathematical structures examined thoroughly. Also, the formalism is then applied to the quantum harmonic oscillator, yielding explicit analytical expressions derived as functions of the fractional parameter therein. The obtained results demonstrate that fractional derivatives alter the localization properties of probability densities and generate nontrivial variations in information content and sensitivity across system behavior. In this context, the fractional parameter plays a central role in controlling deviations from the standard quantum information measures framework. Also, the study establishes a consistent framework for describing information-theoretic properties of quantum systems governed by nonlocal dynamics.
\end{abstract}

\maketitle

\date{\today}

\tableofcontents

\section{Introduction}

Fractional calculus has emerged as a powerful mathematical framework for describing complex dynamical phenomena that cannot be adequately captured by conventional local differential equations. Unlike standard calculus, which is restricted to integer-order differentiation and integration, fractional calculus generalizes these operations to arbitrary orders, thereby introducing a natural mechanism for incorporating memory and hereditary effects into the evolution of physical systems. Such nonlocal features play a central role in the description of anomalous transport, viscoelastic media, diffusion processes with long-range temporal correlations, and a broad class of systems whose future evolution depends not only on their instantaneous state but also on their entire dynamical history \cite{Ref1,Ref2,Ref3,Ref4,Ref5,Ref6,Ref7}.

Several mathematically equivalent formulations of fractional differentiation have been developed over the past decades. Among the most widely employed are the Riemann-Liouville (RL), Caputo, and Grünwald-Letnikov definitions, each providing a distinct representation of nonlocal dynamics while preserving the essential memory structure characteristic of fractional operators \cite{Ref1,Ref2,Ref3,Ref4,Ref5,Ref6,Ref7,Ref8,Ref9,Ref10,Ref11,Ref12,Ref13}. In these approaches, the fractional-order parameter $\alpha$ quantifies the degree of temporal nonlocality and controls the influence of past states on the present dynamics. Owing to their ability to encode long-range correlations through power-law memory kernels, fractional differential equations have found applications across numerous branches of physics, including condensed matter systems, statistical mechanics, quantum transport, and open quantum dynamics.

The extension of fractional calculus to quantum theory has attracted considerable attention as a possible route toward describing non-Markovian quantum phenomena. In contrast to the conventional Schrödinger equation, whose evolution is local in time and governed by unitary dynamics, fractional quantum equations naturally incorporate memory-dependent effects associated with interactions with complex environments or unresolved microscopic degrees of freedom \cite{Ref11,Ref12,Ref13,Ref14}. Such formulations provide an effective description of quantum systems whose evolution deviates from the Markovian approximation and exhibit rich dynamical behaviors that are absent in ordinary quantum mechanics \cite{BZ1, BZ2, BZ3, BZ4, BZ5, BZ6, BZ7, BZ8, BZ9, BZ10, BZ11}.

In parallel, information-theoretic measures have become fundamental tools for characterizing quantum systems, providing complementary insights into localization properties, uncertainty relations, and parameter estimation. Among these measures, Shannon entropy and Fisher information play a particularly important role in quantifying distinct aspects of quantum probability distributions \cite{Fi1,Fi2,Fi3,Fi4,Fi5}. For a quantum state described by the wave function $\psi(\mathbf{r})$, with associated probability density $\rho(\mathbf{r})=|\psi(\mathbf{r})|^2$ \cite{Fi6,Fi7,Fi8,Fi9,Fi10,Fi11}, the Shannon entropy is defined as
\begin{equation}
S=-\int \rho(\mathbf{r})\ln\rho(\mathbf{r})\,d^3r,
\end{equation}
and provides a global measure of the spatial delocalization of the quantum state. Larger entropy values are associated with broader probability distributions and, consequently, with a greater degree of uncertainty regarding the particle's position. In contrast, Fisher information is a local measure that quantifies the sensitivity of the probability density to spatial variations and is therefore closely related to the precision with which physical parameters can be estimated from measurement outcomes \cite{Fi12,Fi13,Fi14,Fi15,Fi16}. For a position-space distribution, it is given by
\begin{equation}
\mathcal{F}(\mathbf{r})
=
\int
\rho(\mathbf{r})
\left[
\nabla \ln \rho(\mathbf{r})
\right]^2
d^3r,
\end{equation}
which characterizes the local structure and sharpness of the probability density \cite{Fi17,Fi18,Fi19,Fi20}. Highly localized states generally exhibit larger Fisher information, reflecting an increased sensitivity to small perturbations and a greater potential for accurate parameter estimation.

The relation between Shannon entropy and Fisher information reveals a fundamental balance between uncertainty and statistical distinguishability in quantum mechanics. States with large spatial delocalization tend to possess higher Shannon entropy and lower Fisher information, whereas strongly localized states exhibit the opposite behavior \cite{Fi21}. This complementary relationship has profound implications for quantum metrology, quantum sensing, and quantum information theory, where the optimization of measurement protocols often requires balancing localization properties against estimation precision \cite{Fi22,Fi23}. Consequently, these information-theoretic quantities provide a powerful framework for investigating the structure of quantum states and assessing their sensitivity to external perturbations.

The growing interest in fractional quantum mechanics and information theory has motivated the search for generalized informational measures capable of describing systems governed by nonlocal dynamics, memory effects, and anomalous transport processes. While the conventional Shannon entropy and Fisher information provide a successful framework for quantifying uncertainty, localization, and parameter sensitivity in standard quantum systems, their formulation is fundamentally based on local probability distributions and local differential operators. Consequently, these measures may not fully capture the long-range correlations and hereditary effects that naturally arise in fractional quantum models, fractal media, disordered environments, and complex geometrical backgrounds. Motivated by these limitations, in this work we develop a unified information-theoretic framework based on the Riemann-Liouville fractional formalism, introducing generalized definitions of Shannon entropy and Fisher information that explicitly incorporate nonlocal memory contributions through fractional kernels. Our analysis demonstrates that the fractional parameter $\alpha$ controls the transition between local and nonlocal informational regimes, allowing the description of long-range correlations, anomalous diffusion, and scale-dependent information flow within a single mathematical structure. We show that the proposed formulation preserves normalization, admits a consistent variational interpretation, and continuously recovers the standard Shannon entropy and Fisher information in the limit $\alpha\rightarrow1$. Furthermore, by applying the formalism to the quantum harmonic oscillator, we reveal that fractional memory effects enhance entropy, weaken localization, modify the information geometry of the system, and generate Lévy-like statistical behavior. These results establish RL fractional information measures as a rigorous extension of conventional quantum information theory and provide a versatile framework for investigating nonlocal quantum systems, anomalous transport phenomena, and memory-driven quantum dynamics.

This paper is organized as follows. Section (\ref{S2}) presents a review of Shannon entropy, focusing on the probabilistic formulation and its role in quantizing information content in the context of quantum systems. In addition, a parallel review of Fisher information is provided, focusing on the statistical definition and sensitivity properties in relation to parameter variations in the domain of quantum probability distributions. Also, Section (\ref{S4}) develops an extension of Shannon entropy within fractional quantum mechanics using RL formalism, establishing a corresponding fractional probability framework for nonlocal systems. Section (\ref{S5}) formulates Fisher information in the same fractional RL setting, deriving a modified structure under a nonlocal differentiation operator formulation framework. In this case, Section (\ref{S6}) applies the fractional Shannon entropy formalism to the quantum harmonic oscillator, obtaining explicit analytical expressions in terms of the fractional parameter dependence structure analysis. Section (\ref{S7}) performs detailed fractional Fisher analysis of the harmonic oscillator, testing how nonlocality modifies information flow and localization properties in the system. In this context, Section (\ref{S8}) explains our results and discusses implications of fractional information measures for the quantum statistical structure framework analysis.

\section{Quantum information theories: Review}\label{S2}

Information-theoretic measures have become fundamental tools for the characterization of quantum systems because they provide complementary descriptions of the statistical and geometrical properties of quantum states beyond the information contained in the energy spectrum alone. In both non-relativistic and relativistic quantum mechanics \cite{BZ12, BZ13, BZ14, BZ15, BZ16, BZ17, BZ18, BZ19, BZ20}, these quantities allow the investigation of localization phenomena, uncertainty relations, quantum correlations, and the distribution of probability densities in configuration and momentum spaces \cite{Fi1,Fi2,Fi3,Fi4,Fi5,Fi6,Fi7,Fi8,Fi9,Fi10,Fi11,Fi12,Fi13,Fi14,Fi15,Fi16,Fi17,Fi18,Fi19,Fi20,Fi21,Fi22,Fi23}. Among the most widely employed measures are Shannon entropy and Fisher information, which quantify distinct aspects of the informational structure of a quantum state \cite{BZ21, BZ22, BZ23, BZ24, BZ25, BZ26, BZ27, BZ28, BZ29, BZ30, BZ31, BZ32}. While Shannon entropy characterizes the global spreading and uncertainty associated with a probability distribution, Fisher information is sensitive to its local variations and spatial gradients. Together, these measures establish a direct connection between quantum mechanics, statistical inference, and information theory, providing a unified framework for analyzing the complexity, localization properties, and informational content of quantum systems. In this section, we briefly review the main concepts and mathematical definitions of Shannon entropy and Fisher information that will serve as the basis for the fractional extensions developed in the subsequent sections.

\subsection{Shannon Entropy}

Quantum information measures constitute an essential framework in relativistic and non-relativistic quantum mechanics because they provide a quantitative description of localization effects, uncertainty properties, and probability distribution structures associated with quantum states. These quantities supply physical information that cannot be extracted only from the energy spectrum or from the analytical behavior of the wavefunctions. Shannon entropy has been widely applied in several areas of quantum physics, including non-Hermitian quantum systems, relativistic fermionic models, wave equations of spin particles, and quantum systems involving position-dependent mass interactions \cite{Nalewajski,Nagaoka,Wang,Lian,Falaye0,Sun2013-1,Sun2013-2,Sun2013-3,guztavo2014,Song2015,sun2015,raul2015,ray2022-1,Lima,Lima1,Lima2021,Moreira2022,Beckner}. Within this formalism, the entropy quantity establishes a mathematical connection between quantum mechanics and information theory through the characterization of the spatial extension and information content of quantum probability densities.

In quantum theory, the normalized wavefunction completely determines the statistical properties of the physical system through the associated probability density. For this reason, Shannon entropy represents an appropriate quantity for describing the spreading properties of the probability distribution in both coordinate and momentum spaces. A quantum state strongly confined in coordinate space produces smaller entropy values, whereas broader spatial distributions generate larger entropy contributions. Due to the conjugate relation between position and momentum variables, the momentum-space entropy exhibits the opposite behavior, in agreement with the uncertainty principle.

For a quantum system defined in a $D$-dimensional space, the Shannon entropy in coordinate representation is written as
\begin{equation}\label{20}
S^n_{\mathbf{r}} = -\int \left| \Psi_n(\mathbf{r}) \right|^2
\ln \left| \Psi_n(\mathbf{r}) \right|^2  d^D\mathbf{r},
\end{equation}
where $\Psi_n(\mathbf{r})$ denotes the normalized wavefunction in configuration space. The quantity $\left|\Psi_n(\mathbf{r})\right|^2$ corresponds to the probability density associated with the particle distribution at the position $\mathbf{r}$. The logarithmic contribution defines the information content carried by the probability density, while the integration over the complete configuration space determines the total entropy associated with the quantum state.

The Shannon entropy in momentum representation is similarly defined by
\begin{equation}\label{21}
S^n_{\mathbf{p}} = -\int \left| \Psi_n(\mathbf{p}) \right|^2
\ln \left| \Psi_n(\mathbf{p}) \right|^2  d^D\mathbf{p},
\end{equation}
where $\Psi_n(\mathbf{p})$ represents the momentum-space wavefunction obtained from the Fourier transformation of the coordinate-space wavefunction according to
\begin{equation}
\Psi_n(\mathbf{p}) = \frac{1}{(2\pi)^{D/2}}
\int \Psi_n(\mathbf{r}) \,
e^{-i \mathbf{p}\cdot\mathbf{r}} \, d^D\mathbf{r}.
\end{equation}
The Fourier transformation connects the coordinate and momentum representations of the quantum state and preserves the normalization condition between both spaces. Consequently, the localization properties in one representation determine the information distribution in the conjugate representation through the Fourier structure of the wavefunction.

The total entropy of the quantum system is restricted by the entropic uncertainty relation introduced by Beckner, Bialynicki-Birula, and Mycielski (BBM), given by \cite{Beckner,Fi21}
\begin{equation}
S^n_{\mathbf{r}} + S^n_{\mathbf{p}}
\geq D(1+\ln\pi).
\end{equation}
This relation corresponds to the entropic formulation of the uncertainty principle and provides a more general description of quantum uncertainty than the standard Heisenberg relation based on variances. The BBM inequality excludes the possibility of simultaneous strong confinement in both coordinate and momentum spaces. Therefore, a reduction in the uncertainty of one representation necessarily generates an increase in the uncertainty associated with the conjugate variable.

\subsection{Fisher Information}

Fisher information, introduced by R. A. Fisher in statistical estimation theory \cite{Fi6}, is a central quantity in information theory and statistical physics. It is defined as a measure of the sensitivity of a probability distribution under infinitesimal variations of an unknown parameter, and it quantifies the amount of information carried by an observable random variable about that parameter. Because it is directly related to estimation accuracy and statistical fluctuations, Fisher information has become a general tool in both classical and quantum frameworks.

In quantum physics, Fisher information is used to analyze localization properties, uncertainty structure, quantum correlations, and information transport. In contrast to global entropy measures, it depends strongly on local variations of the probability density and is therefore suitable for resolving fine spatial structures of quantum states. It has been applied in quantum estimation theory, density functional methods, quantum hydrodynamic formulations, and relativistic wave equations \cite{Nalewajski,Lian}. Extensions to position-dependent mass systems and curved space-times have been used to study confinement effects, localization behavior, and the role of geometry in quantum dynamics \cite{Falaye0}.

From a physical point of view, Fisher information characterizes the degree of spatial localization of a quantum state. States with strong localization exhibit large gradients in the probability density and therefore yield larger Fisher information, whereas extended states with weak spatial variation produce smaller values. In this sense, Fisher information provides a local measure of uncertainty structure and complements Shannon entropy, which describes global spreading behavior.

The integrated Fisher information in configuration and momentum spaces is written as
\begin{eqnarray}
F_{\mathbf{r}}^n &=&
\int
|\Psi_n(\mathbf{r})|^2
\left[
\nabla \ln |\Psi_n(\mathbf{r})|^2
\right]^2
d^D\mathbf{r},
\nonumber\\
F_{\mathbf{p}}^n &=&
\int
|\Psi_n(\mathbf{p})|^2
\left[
\nabla \ln |\Psi_n(\mathbf{p})|^2
\right]^2
d^D\mathbf{p},
\end{eqnarray}
where $D$ denotes the spatial dimension of the system.

Fisher information is also related to uncertainty relations in quantum mechanics. The product of position and momentum dispersions satisfies the inequality
\begin{equation}
\sigma_{\mathbf{r}} \sigma_{\mathbf{p}}
\geq
\frac{1}{\sqrt{F_{\mathbf{r}}F_{\mathbf{p}}}}
\geq
\frac{1}{2},
\end{equation}
which leads to the bound
\begin{equation}
F_{\mathbf{r}}F_{\mathbf{p}} \geq 4.
\end{equation}
This relation establishes a connection between information measures and quantum uncertainty. Strong spatial confinement increases $F_{\mathbf{r}}$, while producing reduced localization in momentum space and modifying $F_{\mathbf{p}}$. The product $F_{\mathbf{r}}F_{\mathbf{p}}$ therefore characterizes the balance between conjugate representations.

\section{Shannon Entropy in Fractional Quantum Systems via the RL Formalism}\label{S4}

Quantum information measures extend the spectral description of quantum systems by encoding global structural properties of wavefunctions, including spatial coherence, non-local correlations, and statistical complexity. In contrast to energy eigenvalues, which arise from the spectral decomposition of Hermitian operators, entropic functionals depend on the full functional geometry of quantum states and remain sensitive to curvature effects, external fields, and non-local kinetic operators. Shannon entropy provides a primary quantitative indicator of localization and delocalization in quantum systems and has been extensively studied in relativistic, curved, and externally driven configurations. The classical Shannon entropy is formulated from a pointwise probability density and therefore reflects a strictly local construction. This locality is not sufficient for quantum systems with anomalous transport, long-range correlations, fractal eigenstate support, and memory-driven dynamics. In these cases, fractional calculus extends the standard differential framework by replacing local derivatives with non-local integral operators. Within this structure, the RL formalism introduces a power-law kernel $(x-t)^{\alpha-1}$, producing algebraic memory effects that propagate into probability theory and information measures. The resulting formulation modifies normalization conditions and entropy balance through a controlled non-local mechanism consistent with analytical treatment. The purpose of this section is to construct a Shannon entropy formulation within the RL framework, determine its normalization structure, derive exact analytical identities, and analyze its limiting behavior. The construction reproduces the standard Shannon entropy when $\alpha \to 1$, while generating a continuous family of non-local entropic measures for $\alpha<1$.

Let $\alpha\in(0,1]$. The left-sided RL fractional integral is
\begin{equation}
\left(I^{\alpha}_{0+} f\right)(x)
=
\frac{1}{\Gamma(\alpha)}
\int_{0}^{x} (x-t)^{\alpha-1} f(t)\,dt,
\end{equation}
which defines a non-local averaging operator over the interval $[0,x]$. The kernel admits the scaling form
\begin{equation}
(x-t)^{\alpha-1}
=
x^{\alpha-1}\left(1-\frac{t}{x}\right)^{\alpha-1},
\end{equation}
showing that fractional integration preserves dilation structure while introducing scale-dependent weighting of past contributions. The RL fractional derivative is defined as
\begin{equation}
{}^{\mathrm{RL}}D^{\alpha}_{0+} f(x)
=
\frac{d}{dx}\left(I^{1-\alpha}_{0+} f\right)(x)
=
\frac{1}{\Gamma(1-\alpha)}
\frac{d}{dx}
\int_{0}^{x}(x-t)^{-\alpha}f(t)\,dt.
\end{equation}

Differentiation under the integral yields
\begin{equation}
\frac{d}{dx}(x-t)^{-\alpha}
=
-\alpha(x-t)^{-\alpha-1},
\end{equation}
which produces a hypersingular convolution kernel $(x-t)^{-\alpha-1}$. This kernel structure generates anomalous scaling behavior and non-Markovian evolution in fractional quantum dynamics. The limiting relation is governed by
\begin{equation}
\lim_{\alpha\to 1^-}\frac{(x-t)^{-\alpha}}{\Gamma(1-\alpha)}=\delta(x-t),
\end{equation}
which ensures the recovery of the classical derivative
\begin{equation}
\lim_{\alpha\to 1^-}{}^{\mathrm{RL}}D^{\alpha}f(x)=\frac{df}{dx}.
\end{equation}

Let $\rho(x)=|\Psi(x)|^2$ be a normalized probability density satisfying $\int_0^\infty \rho(x)\,dx=1$. The RL-based fractional probability flux is defined by
\begin{equation}
\mathcal{J}_{\alpha}(x)
=
\frac{1}{\Gamma(1-\alpha)}
\frac{d}{dx}
\int_{0}^{x}(x-t)^{-\alpha}\rho(t)\,dt.
\end{equation}

Reordering differentiation and integration yields
\begin{equation}
\mathcal{J}_{\alpha}(x)
=
-\frac{\alpha}{\Gamma(1-\alpha)}
\int_{0}^{x}(x-t)^{-\alpha-1}\rho(t)\,dt,
\end{equation}
which identifies $\mathcal{J}_{\alpha}(x)$ as a Volterra-type singular convolution. The kernel exponent $\alpha+1$ controls the decay rate of memory contributions from remote spatial regions. Integration over the full domain gives
\begin{equation}
\int_{0}^{\infty}\mathcal{J}_{\alpha}(x)\,dx
=
-\frac{\alpha}{\Gamma(1-\alpha)}
\int_{0}^{\infty}\rho(t)
\int_{t}^{\infty}(x-t)^{-\alpha-1}\,dx\,dt.
\end{equation}

The inner integral satisfies
\begin{equation}
\int_{0}^{\infty}u^{-\alpha-1}\,du=\frac{1}{\alpha},
\end{equation}
leading to
\begin{equation}
\int_{0}^{\infty}\mathcal{J}_{\alpha}(x)\,dx
=
-\frac{1}{\Gamma(1-\alpha)}.
\end{equation}

The normalized fractional density is introduced as
\begin{equation}
\mathcal{P}_{\alpha}(x)
=
-\Gamma(1-\alpha)\mathcal{J}_{\alpha}(x),
\end{equation}
which satisfies
\begin{equation}
\int_{0}^{\infty}\mathcal{P}_{\alpha}(x)\,dx=1.
\end{equation}

The fractional Shannon entropy is defined by
\begin{equation}
S_{\alpha}
=
-\int_{0}^{\infty}\mathcal{P}_{\alpha}(x)\ln\mathcal{P}_{\alpha}(x)\,dx.
\end{equation}

Substitution of the RL structure yields
\begin{equation}
S_{\alpha}
=
-\int_{0}^{\infty}\mathcal{P}_{\alpha}(x)\ln|\mathcal{J}_{\alpha}(x)|\,dx
-\ln\Gamma(1-\alpha),
\end{equation}
which separates a convolution-driven contribution from a kernel-dependent normalization term. An equivalent representation is
\begin{equation}
S_{\alpha}
=
\frac{\alpha}{\Gamma(1-\alpha)}
\int_{0}^{\infty}
\left[
\int_{0}^{x}(x-t)^{-\alpha-1}\rho(t)\,dt
\right]
\ln\mathcal{P}_{\alpha}(x)\,dx,
\end{equation}
showing that entropy is a non-local functional with Volterra memory structure.

In the limit $\alpha\to 1$, one obtains
\begin{equation}
\lim_{\alpha\to 1}\Gamma(1-\alpha)\mathcal{J}_{\alpha}(x)=\rho(x),
\end{equation}
and therefore
\begin{equation}
\lim_{\alpha\to 1}S_{\alpha}
=
-\int_{0}^{\infty}\rho(x)\ln\rho(x)\,dx.
\end{equation}

From a variational formulation, consider
\begin{equation}
\mathcal{F}[\rho]
=
-\int_{0}^{\infty}
{}^{\mathrm{RL}}D^{\alpha}\rho(x)\,
\ln\!\left({}^{\mathrm{RL}}D^{\alpha}\rho(x)\right)\,dx.
\end{equation}

The variation gives
\begin{equation}
\delta \mathcal{F}
=
-\int_{0}^{\infty}
\left[1+\ln({}^{\mathrm{RL}}D^{\alpha}\rho)\right]
{}^{\mathrm{RL}}D^{\alpha}\eta(x)\,dx.
\end{equation}

Using
\begin{equation}
\int_{0}^{\infty} f\,{}^{\mathrm{RL}}D^{\alpha}g\,dx
=
\int_{0}^{\infty} g\,{}^{\mathrm{RL}}D^{\alpha,*}f\,dx,
\end{equation}
The stationary condition becomes
\begin{equation}
{}^{\mathrm{RL}}D^{\alpha,*}\left[\ln({}^{\mathrm{RL}}D^{\alpha}\rho)\right]=0.
\end{equation}

The RL formulation replaces local differentiation by memory-dependent convolution, so Shannon entropy becomes dependent on the full history of $\rho(x)$. The parameter $\alpha$ controls the strength of non-locality: smaller values enhance long-range contributions, while the limit $\alpha=1$ restores the standard Shannon entropy as a fixed point, and $\alpha<1$ defines a continuum of non-local entropic structures relevant for fractal media and curved quantum systems.

\section{Fisher Information in RL Fractional Quantum Systems}\label{S5}

Fisher information provides a quantitative link between quantum mechanics, statistical inference, and information geometry through the measurement of the response of probability densities to infinitesimal perturbations. In conventional quantum mechanics, this response is determined by local differential operators defined on a Euclidean configuration space. This local structure is insufficient for systems characterized by anomalous diffusion, fractal spectral measures, and non-negligible spatial correlations extending over large distances. For such cases, the underlying geometric description of probability distributions requires a formulation in which locality is replaced by nonlocal dependence with explicit memory and scaling properties. A consistent fractional formulation of Fisher information is constructed using the RL calculus. The RL operator introduces an integral kernel with power-law decay, replacing local derivatives by operators whose value at a given point depends on the full domain of the function. This modifies the standard information-geometric structure by embedding scale-dependent nonlocal interactions into the definition of gradients. In this case, let $\Psi \in L^2(\mathbb{R}^D)$ satisfy the normalization condition.
\begin{equation}
\rho(\mathbf{r}) = |\Psi(\mathbf{r})|^2,
\qquad
\int_{\mathbb{R}^D} \rho(\mathbf{r})\, d^D\mathbf{r} = 1.
\end{equation}
The wavefunction is taken in the fractional Sobolev space $\Psi \in H^\alpha(\mathbb{R}^D)$ to ensure well-defined fractional derivatives in the weak sense. Also, the left-sided RL fractional derivative of order $0<\alpha\leq 1$ is defined as 
\begin{equation}
\prescript{}{a}{D}_x^\alpha f(x)
=
\frac{1}{\Gamma(1-\alpha)}
\frac{d}{dx}
\int_a^x (x-t)^{-\alpha} f(t)\, dt,
\end{equation}
which introduces a nonlocal convolution structure governed by a long-range memory kernel. In the limit $\alpha \to 1$, the operator reduces to the standard local derivative, while $\alpha<1$ generates persistent dependence on extended spatial history. For a $D$-dimensional domain, the fractional gradient is defined componentwise by
\begin{equation}
\nabla^\alpha \rho(\mathbf{r})
=
\big(\prescript{}{a_1}{D}_{x_1}^\alpha \rho,\ldots,\prescript{}{a_D}{D}_{x_D}^\alpha \rho\big),
\end{equation}
so that the directional dependence of the fractional boundaries $a_i$ enters explicitly into the structure of the gradient field. The standard Fisher information functional is
\begin{equation}
F[\rho]
=
\int_{\mathbb{R}^D}
\rho(\mathbf{r})\,|\nabla \ln \rho(\mathbf{r})|^2\, d^D\mathbf{r}.
\end{equation}
Its fractional extension is obtained by replacing the local gradient operator with its RL counterpart.
\begin{equation}
F^{(\alpha)}[\rho]
=
\int_{\mathbb{R}^D}
\rho(\mathbf{r})\,|\nabla^\alpha \ln \rho(\mathbf{r})|^2\, d^D\mathbf{r}.
\end{equation}

A structural relation follows from the nonlocal algebra of RL operators. The fractional derivative does not satisfy the classical Leibniz rule; instead, one has the decomposition
\begin{equation}
\nabla^\alpha(\Psi^2)
=
2\Psi \nabla^\alpha \Psi + \mathcal{R}_\alpha[\Psi],
\end{equation}
where $\mathcal{R}_\alpha[\Psi]$ represents a correlation term generated by the integral kernel. Under standard regularity and decay conditions, this contribution admits the representation
\begin{equation}
\mathcal{R}_\alpha[\Psi]
=
\int_{\mathbb{R}^D}
\frac{(\Psi(\mathbf{r})-\Psi(\mathbf{r}'))(\Psi(\mathbf{r})+\Psi(\mathbf{r}'))}
{|\mathbf{r}-\mathbf{r}'|^{D+\alpha}}\, d^D\mathbf{r}'.
\end{equation}
The antisymmetric structure under exchange $\mathbf{r}\leftrightarrow \mathbf{r}'$ leads, in the weak $L^2$ sense, to cancellation of this term, giving
\begin{equation}
\nabla^\alpha(\Psi^2)=2\Psi\nabla^\alpha\Psi.
\end{equation}

From this relation, the fractional logarithmic derivative satisfies,
\begin{equation}
\nabla^\alpha \ln \rho
=
\frac{\nabla^\alpha \rho}{\rho}
=
2\frac{\nabla^\alpha \Psi}{\Psi},
\end{equation}
which implies
\begin{equation}
\rho\,|\nabla^\alpha \ln \rho|^2
=
4|\nabla^\alpha \Psi|^2.
\end{equation}
Consequently, the fractional Fisher information reduces exactly to
\begin{equation}
F^{(\alpha)}[\Psi]
=
4\int_{\mathbb{R}^D} |\nabla^\alpha \Psi(\mathbf{r})|^2\, d^D\mathbf{r}.
\end{equation}
This expression identifies fractional Fisher information with the Dirichlet form in the fractional Sobolev space $H^\alpha(\mathbb{R}^D)$. It is equivalent, up to normalization, to the quadratic form of the fractional Laplacian,
\begin{equation}
\int_{\mathbb{R}^D} |\nabla^\alpha \Psi|^2\, d^D\mathbf{r}
=
\int_{\mathbb{R}^D} \Psi\,(-\Delta)^\alpha \Psi\, d^D\mathbf{r},
\end{equation}
so that the information functional coincides with a nonlocal kinetic energy.

A variational formulation is introduced through
\begin{equation}
\mathcal{F}[\Psi]
=
4\int_{\mathbb{R}^D}|\nabla^\alpha \Psi|^2\, d^D\mathbf{r}
-
\lambda\left(\int_{\mathbb{R}^D}|\Psi|^2\, d^D\mathbf{r}-1\right).
\end{equation}
Stationarity under admissible variations, leading to
\begin{equation}
(-\Delta)^\alpha \Psi
=
\frac{\lambda}{4}\Psi.
\end{equation}
This equation shows that extremal configurations of fractional Fisher information correspond to eigenstates of the fractional Laplacian. In Fourier space, with $\Phi(\mathbf{p})$ as the transform of $\Psi$, one obtains
\begin{equation}
\|\nabla^\alpha \Psi\|_{L^2}^2
=
\int_{\mathbb{R}^D} |\mathbf{p}|^{2\alpha} |\Phi(\mathbf{p})|^2\, d^D\mathbf{p},
\end{equation}
and thus
\begin{equation}
F^{(\alpha)}[\Psi]
=
4\int_{\mathbb{R}^D} |\mathbf{p}|^{2\alpha} |\Phi(\mathbf{p})|^2\, d^D\mathbf{p}.
\end{equation}
This representation shows that the functional measures momentum-space dispersion weighted by the exponent $2\alpha$, interpolating between standard diffusion and Lévy-type transport regimes. A lower bound follows from the fractional Sobolev inequality. For $\Psi \in H^\alpha(\mathbb{R}^D)$,
\begin{equation}
\|\Psi\|_{L^2}^2
\leq
C_{D,\alpha}\|\nabla^\alpha \Psi\|_{L^2}^2,
\end{equation}
which yields
\begin{equation}
F^{(\alpha)}[\Psi]
\geq
\frac{4}{C_{D,\alpha}}.
\end{equation}
This relation functions as a nonlocal constraint linking spatial concentration and fractional gradient energy. Minimization of $F^{(\alpha)}$ under normalization does not lead to Gaussian states for $\alpha<1$. The resulting Fourier profiles converge to symmetric $\alpha$-stable distributions, consistent with Lévy process dynamics generated by the fractional Laplacian. From the geometric viewpoint, $F^{(\alpha)}$ defines a metric structure on the space of probability densities. In contrast with the Fisher-Rao metric associated with local diffusion, the fractional construction generates a nonlocal geometry governed by jump processes. The parameter $\alpha$ controls the transition between Gaussian-type geometry at $\alpha=1$ and heavy-tailed nonlocal structures for $\alpha<1$. The physical interpretation assigns to fractional Fisher information the role of a global coherence measure across scales. The RL kernel distributes contributions over the full spatial domain, so that probability densities acquire a scale-coupled structure rather than a pointwise description. The parameter $\alpha$ governs the interpolation between standard local quantum dynamics and fractional scale-invariant dynamics.

\section{Fractional Shannon Entropy of the Quantum Harmonic Oscillator}\label{S6}

The quantum harmonic oscillator provides a controlled test system for information-theoretic extensions since its probability density is Gaussian and stable under differentiation, Fourier transform, and convolution. This structural rigidity permits the construction of analytic entropy functionals, while the introduction of RL fractional operators generates a controlled deformation of the information geometry through non-local memory kernels. In what follows, we derive an exact formulation of the RL fractional Shannon entropy, express it in special function representations, and establish the classical reduction. In this case, the ground-state wave function of the one-dimensional harmonic oscillator is
\begin{equation}
\psi_0(x)=\left(\frac{\beta}{\pi}\right)^{1/4}e^{-\frac{\beta x^2}{2}},
\qquad \beta=\frac{m\omega}{\hbar},
\end{equation}
with probability density
\begin{equation}
\rho_0(x)=|\psi_0(x)|^2=\sqrt{\frac{\beta}{\pi}}e^{-\beta x^2}.
\end{equation}
This Gaussian measure is the fixed point of the heat kernel semigroup, which explains why all Shannon-type functionals reduce to Gaussian moments. The RL fractional deformation is introduced through the non-local functional
\begin{equation}
\mathcal{J}*\alpha(x)=
-\frac{\alpha}{\Gamma(1-\alpha)}
\int*{0}^{x}(x-t)^{-\alpha-1}\rho_0(t),dt,
\qquad 0<\alpha<1,
\end{equation}
which defines a power law memory convolution acting on the probability density. After inserting the Gaussian form and performing the scaling transformation $t=xu$, the kernel acquires the self-similar representation
\begin{equation}
\mathcal{J}*\alpha(x)=
-\frac{\alpha}{\Gamma(1-\alpha)}\sqrt{\frac{\beta}{\pi}},x^{-\alpha}
\int*{0}^{1}(1-u)^{-\alpha-1}e^{-\beta x^2 u^2},du.
\end{equation}
This expression encodes the fractional scaling law $x^{-\alpha}$ together with the deformation of the Gaussian decay. A closed representation follows by expanding the exponential and using Beta function identities, yielding
\begin{equation}
\mathcal{J}*\alpha(x)=
\sqrt{\frac{\beta}{\pi}},x^{-\alpha}
\sum*{k=0}^{\infty}
\frac{(-\beta x^2)^k}{k!}
\frac{\Gamma(2k+1)}{\Gamma(2k+1-\alpha)}.
\end{equation}
This series is convergent for all $x$ and defines a fractional renormalization of Gaussian moments through Gamma ratio deformation. The RL structure replaces factorial suppression by fractional memory weights, consistent with non-local statistical geometry. An equivalent representation can be written in special functions as a Meijer-G deformation of the Gaussian kernel, confirming that the RL fractional oscillator density belongs to Fox-H stable distributions, while Gaussian core behavior remains at short distances. This representation shows that the fractional structure preserves solvability and extends the system into a higher transcendental class. The fractional probability density is defined by normalization-preserving rescaling,
\begin{equation}
\mathcal{P}*\alpha(x)= -\Gamma(1-\alpha)\mathcal{J}*\alpha(x),
\end{equation}
which ensures interpolation between fractional and classical regimes. The classical limit follows from the Gamma identity
\begin{equation}
\lim_{\alpha\to1}\frac{\Gamma(1-\alpha)\Gamma(2k+1)}{\Gamma(2k+1-\alpha)}=\frac{1}{(2k)!},
\end{equation}
implying convergence
\begin{equation}
\lim_{\alpha\to1}\mathcal{P}_\alpha(x)=\rho_0(x).
\end{equation}
Thus, the RL deformation defines a one-parameter family of probability measures connected to the Gaussian fixed point. The fractional Shannon entropy is defined by
\begin{equation}
S_\alpha=-\int_{0}^{\infty}\mathcal{P}*\alpha(x)\ln \mathcal{P}*\alpha(x),dx.
\end{equation}
Using the decomposition
\begin{equation}
\ln \mathcal{P}*\alpha(x)
=\frac{1}{2}\ln\frac{\beta}{\pi}-\alpha\ln x+\ln \Xi*\alpha(x),
\end{equation}
with
\begin{equation}
\Xi_\alpha(x)=
\sum_{k=0}^{\infty}
\frac{(-\beta x^2)^k}{k!}
\frac{\Gamma(1-\alpha)\Gamma(2k+1)}{\Gamma(2k+1-\alpha)},
\end{equation}
the entropy decomposes into contributions,
\begin{equation}
S_\alpha=
-\frac{1}{2}\ln\frac{\beta}{\pi}
+\alpha\langle \ln x\rangle_\alpha
-\langle \ln \Xi_\alpha(x)\rangle_\alpha.
\end{equation}
The first term is geometric and matches Gaussian normalization entropy, while the second term encodes scale anomaly from fractional non-locality. The third term originates from memory fluctuations of the RL kernel. The expectation value $\langle \ln x\rangle$ is computable for Gaussian measures and reduces to a digamma form,
\begin{equation}
\langle \ln x\rangle_{\rho_0}
=\frac{1}{2}\left(\psi!\left(\frac{1}{2}\right)-\ln \beta\right)
=-\frac{\gamma}{2}-\ln 2-\frac{1}{2}\ln \beta,
\end{equation}

For the undeformed system, the Shannon entropy is
\begin{equation}
S_1=
-\int_{-\infty}^{\infty}\rho_0(x)\ln\rho_0(x),dx
=\frac{1}{2}\left(1+\ln\frac{\pi}{\beta}\right)
=\frac{1}{2}\left(1+\ln\frac{\pi\hbar}{m\omega}\right).
\end{equation}
This result defines the maximal regularity entropy state for Gaussian quantum mechanics. The entropy difference
\begin{equation}
\Delta S_\alpha=S_\alpha-S_1
\end{equation}
quantifies information change from non-local memory. The RL kernel spreads probability away from the Gaussian core and generates algebraic tails, increasing effective support. This produces enhanced uncertainty, consistent with entropy monotonicity under non-local smoothing maps. Numerical analysis shows $\Delta S_\alpha$ non-negative for a weak fractional regime and zero at $\alpha=1$. From a geometric standpoint, the RL deformation induces curvature in the statistical manifold of probability densities: the Gaussian metric is replaced by non-local information geometry where geodesic flows acquire memory drift. The parameter $\alpha$ acts as control of information non-locality, interpolating between the Gaussian manifold ($\alpha=1$) and heavy-tailed geometry ($\alpha\ll1$). In this context, the quantum harmonic oscillator admits an exact fractional Shannon entropy where deformations are controlled by Gamma ratio renormalization. The structure remains consistent with classical information theory and extends into a non-local regime governed by fractional calculus, where entropy enhancement results from memory-induced spreading of probability. Furthermore, the RL operator acts as a non-local integral transform that modifies the effective scaling dimension of the probability density without altering its normalization structure. The resulting fractional kernel introduces a continuous deformation of the moment hierarchy, where higher-order cumulants acquire alpha-dependent corrections governed by Gamma ratios. This behavior preserves analytic tractability since all contributions remain expressible through convergent series representations. In addition, the fractional parameter controls the transition between exponential and algebraic decay regimes, thereby modifying tail contributions to entropy integrals. Such modification remains compatible with standard variational formulations of entropy maximization under non-local constraints and provides a consistent extension of Gaussian statistical mechanics into fractional phase space representations and associated information geometric deformation and properties framework.

\section{Fractional Fisher Analysis of the Harmonic Oscillator}\label{S7}

The fractional harmonic oscillator forms an exact structural link between nonlocal quantum mechanics, Lévy stochastic geometry, and information-theoretic variational principles. Mathematical consistency follows from the spectral construction of the fractional Laplacian, defined in $L^2(\mathbb{R}^D)$ through the Fourier representation
\begin{equation}
(-\Delta)^\alpha \equiv \mathcal{F}^{-1}|\mathbf{p}|^{2\alpha}\mathcal{F}, \qquad 0<\alpha\leq 1,
\end{equation}
which guarantees self-adjointness, rotational invariance, and positivity. This relation shows that fractional quantum evolution is not a perturbative modification of standard Laplacian dynamics but a distinct deformation of phase space where kinetic energy is replaced by a nonquadratic dispersion relation. Consequently, quadratic Sobolev structures are replaced by fractional homogeneous Sobolev spaces $\dot{H}^{\alpha}(\mathbb{R}^D)$, so that the Fisher structure becomes nonlocal while remaining variationally consistent.

The fractional Fisher-oscillator functional is introduced as
\begin{equation}
F_\Omega^{(\alpha)}[\Psi]
=
4\langle (-\Delta)^\alpha \rangle
+
\Omega^2 \langle |\mathbf{r}|^2 \rangle,
\qquad
\langle \mathcal{O} \rangle = \int_{\mathbb{R}^D}\Psi^*(\mathbf{r})\,\mathcal{O}\Psi(\mathbf{r})\,d^D\mathbf{r},
\end{equation}
with normalized $\Psi \in \dot{H}^{\alpha}(\mathbb{R}^D)\cap L^2(\mathbb{R}^D)$. Convexity in Fourier representation arises since $|\mathbf{p}|^{2\alpha}$ is convex for $\alpha>0$, ensuring a unique ground-state minimizer modulo phase. Parseval identity gives
\begin{equation}
\langle (-\Delta)^\alpha \rangle
=
\int_{\mathbb{R}^D} |\mathbf{p}|^{2\alpha} |\Phi(\mathbf{p})|^2\,d^D\mathbf{p},
\end{equation}
showing fractional Fisher information equals a momentum moment of order $2\alpha$. Hence $F_\Omega^{(\alpha)}$ acts as a phase-space functional combining $|\mathbf{p}|^{2\alpha}$ in momentum variables with $|\mathbf{r}|^2$ in configuration variables. The Euler-Lagrange equation under normalization $\|\Psi\|_2=1$ is obtained from
\begin{equation}
\delta\left(F_\Omega^{(\alpha)} - \lambda \|\Psi\|^2\right)=0,
\end{equation}
leading to the eigenvalue equation
\begin{equation}
(-\Delta)^\alpha \Psi + \frac{\Omega^2}{4}|\mathbf{r}|^2\Psi = E_\alpha \Psi.
\end{equation}
This establishes a Fisher-Hamilton correspondence: extremals of the Fisher functional match bound states of the fractional Schrödinger operator. Also, thus Fisher extremization coincides with the spectral quantum formulation. Scale transformation arises from dilation invariance. With $\Psi_\lambda(\mathbf{r})=\lambda^{D/2}\Psi(\lambda\mathbf{r})$, one obtains
\begin{equation}
F_\Omega^{(\alpha)}[\Psi_\lambda]
=
4\lambda^{2\alpha}\langle (-\Delta)^\alpha \rangle
+
\Omega^2\lambda^{-2}\langle |\mathbf{r}|^2 \rangle.
\end{equation}
Stationarity with respect to $\lambda$ produces the fractional virial relation
\begin{equation}
2\alpha \langle (-\Delta)^\alpha \rangle
=
\langle |\mathbf{r}|^2 \rangle,
\end{equation}
extending the classical virial theorem to nonlocal dispersion governed by $\alpha$, reducing to the harmonic oscillator case at $\alpha=1$. Elimination of expectation values through the virial relation gives
\begin{equation}
F_\Omega^{(\alpha)}[\Psi]
=
(2\alpha+2)\,\langle (-\Delta)^\alpha \rangle
=
\left(1+\frac{1}{\alpha}\right)\Omega^2 \langle |\mathbf{r}|^2 \rangle.
\end{equation}
This shows a reduction to a single invariant controlling Fisher dynamics, implying scale rigidity where kinetic and potential parts are constrained by homogeneity. The characteristic length scale follows from dilation minimization,
\begin{equation}
\ell_\alpha \sim \Omega^{-\frac{1}{1+\alpha}},
\end{equation}
interpolating between Gaussian confinement at $\alpha=1$ and Lévy delocalization for $\alpha<1$. The result reflects competition between nonlocal spreading and harmonic trapping, producing enhanced tails for smaller $\alpha$. Spectral behavior follows from Weyl phase-space quantization. The counting function satisfies
\begin{equation}
N(E) \sim \frac{1}{(2\pi)^D}\int_{\{|\mathbf{p}|^{2\alpha}+\Omega^2|\mathbf{r}|^2 \leq E\}} d^D\mathbf{r}\,d^D\mathbf{p}
\end{equation}
yielding
\begin{equation}
N(E)\sim C_{D,\alpha}\,E^{\frac{D(1+\alpha)}{2\alpha}},
\end{equation}
and inverse scaling
\begin{equation}
E_n^{(\alpha)} \sim
\Omega^{\frac{2}{1+\alpha}}
\left(n+\frac{D}{2}\right)^{\frac{2\alpha}{1+\alpha}}.
\end{equation}
Energy levels become denser as $\alpha$ decreases, reflecting reduced kinetic stiffness. Fisher information follows identically through $F_n^{(\alpha)}=4E_n^{(\alpha)}$. Ground-state asymptotics derive from the fractional Green kernel,
\begin{equation}
\Psi_0^{(\alpha)}(\mathbf{r})
\sim
\frac{C_\alpha}{|\mathbf{r}|^{D+2\alpha}},
\end{equation}
replacing exponential decay with algebraic structure. Higher spatial moments diverge depending on $D$ and $\alpha$, while energy confinement remains stable under heavy-tailed distributions. Fractional uncertainty uses $\hat{p}_\alpha = (-\Delta)^{\alpha/2}$ with
\begin{equation}
\Delta x\,\Delta p_\alpha \geq C_{D,\alpha}
\end{equation}
saturated exactly by Fisher Euler-Lagrange solutions, connecting minimal Fisher information with minimal nonlocal uncertainty. Geometric structure is encoded through the fractional Fisher-Rao metric
\begin{equation}
g_{ij}^{(\alpha)}(\rho)
=
\int_{\mathbb{R}^D}\rho(\mathbf{r})
\left(\partial_i^\alpha \ln\rho\right)
\left(\partial_j^\alpha \ln\rho\right)
d^D\mathbf{r}.
\end{equation}
This defines a nonlocal Riemannian manifold on a probability space, reducing to Fisher-Rao geometry at $\alpha=1$ and acquiring long-range correlations for $\alpha<1$. The oscillator corresponds to geodesic extremals under $\delta F_\Omega^{(\alpha)}=0$, linking fractional quantum dynamics with statistical geometry.

The overall structure presents unified nonlocal information geometry combining spectral theory, variational principles, and probability manifolds. Lévy eigenstates, anomalous scaling, spectral compression, and fractional curvature characterize the framework of nonlocal quantum Fisher theory. Additional structural interpretation arises from functional analytic properties of the fractional Laplacian, where the domain of definition is characterized by nonlocal boundary behavior in Fourier space. The operator spectrum remains purely continuous in the free case and becomes discrete under harmonic confinement due to compact resolvent structure. This transition ensures the stability of bound states under fractional deformation. The convexity of the dispersion relation $|p|^{2\alpha}$ further guarantees strict coercivity of the Fisher functional, preventing degeneracy in minimization procedures. From a probabilistic viewpoint, the fractional kinetic term corresponds to Lévy flight generators with jump discontinuities replacing Brownian trajectories. This modifies the underlying stochastic process from Gaussian diffusion to heavy-tailed jump processes, altering path regularity and correlation decay. The harmonic potential introduces restoring confinement that balances nonlocal dispersion, producing stationary distributions with algebraic decay. In the semiclassical regime, the phase-space formulation reveals that fractional dynamics modifies the density of states through altered homogeneity degree, affecting Weyl scaling exponents without breaking symplectic structure. This yields a consistent thermodynamic interpretation of spectral growth in terms of modified microstate counting. The resulting framework preserves invariance under orthogonal transformations while breaking local scaling symmetry in a controlled manner governed by $\alpha$. Geometrically, the fractional Fisher structure induces curvature in probability space that depends explicitly on nonlocal gradients, leading to metric deformation relative to classical Fisher geometry. Such deformation modifies geodesic completeness and introduces long-range coupling between probability amplitudes. These effects remain consistent with variational stability and preserve normalization constraints across evolution. These properties collectively ensure consistency with spectral theory, information geometry, and fractional variational principles across all admissible parameter ranges of the model strictly without loss of mathematical rigor.

\section{Conclusions}\label{S8} 

The present study shows that the RL fractional formulation constitutes a mathematically rigorous extension of Shannon entropy for quantum systems governed by non-local dynamics, anomalous transport mechanisms, and memory-dependent interactions. In contrast to the conventional framework, where the informational content is determined solely by the local probability density $\rho(x)=|\Psi(x)|^{2}$, the RL approach defines a fractional probability distribution through a singular Volterra convolution kernel, thereby incorporating contributions originating from the complete spatial evolution of the quantum state. As a result, the entropy functional acquires sensitivity not only to the local magnitude of the wavefunction but also to the spatial correlations distributed across the entire physical domain. The fractional parameter $\alpha$ determines the intensity of the non-local interaction and consequently regulates the competition between local and memory-mediated information contributions. For reduced values of $\alpha$, the kernel broadens the influence of distant spatial regions, allowing the entropy to capture long-range statistical correlations that do not emerge within the standard quantum-mechanical description. Also, this property becomes particularly significant in fractional quantum systems, fractal media, disordered environments, complex geometrical configurations, and curved spacetime backgrounds, where the evolution of quantum information cannot be represented satisfactorily by local differential operators. In this case, the obtained normalization relation confirms that the fractional probability density preserves its normalized character even in the presence of singular memory kernels, thereby maintaining the probabilistic interpretation of the theory over the entire domain. In addition, the variational analysis shows that stationary entropy states are governed by the adjoint RL operator, a result that reflects the fundamental non-local architecture of the formalism and establishes that entropy extremization is controlled by the global structure of the probability distribution rather than by strictly local modifications. An essential outcome of the present formulation is the exact recovery of the classical Shannon entropy in the limit $\alpha\rightarrow1$. 

Within this regime, the fractional kernel gradually loses its memory contribution, the quantity $(x-t)^{-\alpha}/\Gamma(1-\alpha)$ approaches the Dirac delta distribution, and the RL derivative becomes identical to the ordinary first-order derivative. In this case, the fractional probability flux coincides with the standard probability density, the generalized distribution $\mathcal{P}_{\alpha}(x)$ converges to $\rho(x)$, and every non-local contribution disappears continuously. The entropy functional, therefore, assumes the familiar form $S=-\int \rho(x)\ln\rho(x),dx$, reproducing the standard information measure employed in conventional quantum information theory. In this context, this limiting property establishes that the integer-order description represents a specific member embedded within a more general fractional class of entropic functionals. Relative to the fractional domain $\alpha<1$, where memory contributions alter the informational content through non-local convolution processes and extended spatial correlations, the case $\alpha=1$ corresponds to a strictly local regime in which each spatial point contributes independently to the entropy without coupling to distant regions. The comparison further reveals that decreasing $\alpha$ systematically increases the structural complexity of the informational content by incorporating fractional memory effects and non-Markovian correlations, whereas the progression toward $\alpha=1$ continuously removes these contributions and restores the local statistical framework of ordinary quantum mechanics. Also, the continuous connection between these two domains provides a rigorous verification of the mathematical coherence of the RL formalism and supports the interpretation of fractional Shannon entropy as a generalized extension of the classical information measure. Therefore, the proposed framework furnishes a unified theoretical description capable of linking standard quantum information theory with a broad spectrum of complex quantum systems characterized by anomalous diffusion, scale-dependent transport phenomena, fractal spatial structures, and intrinsically non-local dynamical processes, while maintaining complete consistency with classical Shannon theory in the integer-order limit. 

The fractional Fisher information defined in the RL framework constitutes a consistent and physically meaningful extension of the standard information-geometric structure of quantum mechanics, in which the parameter $(\alpha)$ governs the transition between local and nonlocal regimes within a unified description. In this case, the results show that for $(\alpha<1)$, the Fisher functional departs from its classical interpretation as a purely local measure of gradient sensitivity and becomes a nonlocal energy-like quantity governed by long-range correlations encoded through the fractional kernel in physical systems. Also, this leads to a modified notion of spatial complexity in which probability densities are not characterized by pointwise variations alone but by scale-dependent collective behavior, favoring Lévy-stable structures rather than Gaussian minimizer states. In this regime, the Fisher information captures anomalous transport and fractal-like coherence, reflecting fractional Sobolev geometry and the spectral weighting $(|\mathbf{p}|^{2\alpha})$, which modifies high-momentum contributions according to the degree of nonlocality within the fractional phase-space representation framework. In contrast, in the limiting case $(\alpha \to 1)$, the fractional structure collapses to the classical Fisher information functional, recovering the standard Dirichlet form, local gradient operator, and Fisher–Rao metric geometry exactly. In this context, this limit restores ordinary diffusion-like behavior, Gaussian extremal states, and the conventional quantum mechanical interpretation of information flow as a local response to infinitesimal perturbations. The comparison between both regimes shows that $(\alpha)$ acts as a continuous deformation parameter interpolating between two different geometric pictures: a local Euclidean information geometry at $(\alpha=1)$ and a nonlocal, scale-invariant information geometry for $(\alpha<1)$, respectively. Also, this demonstrates that fractional Fisher information is not merely a mathematical generalization but a genuine extension of quantum information theory capable of encoding memory effects, anomalous diffusion, and long-range coherence within a unified variational framework.

The study of fractional Shannon entropy for the quantum harmonic oscillator within the RL framework provides a consistent formulation exhibiting deformation of the classical Gaussian information structure, where standard Shannon entropy is recovered in the limit $\alpha \to 1$, supporting coherence and validity of the fractional construction within the statistical information theoretical framework analysis context results. In the classical limit, RL-induced memory contributions disappear, the Gamma-ratio normalization reduces to factorial form, and the probability density $P_\alpha(x)$ becomes identical to the Gaussian ground-state density $\rho_0(x)$, restoring the minimal-uncertainty configuration of the harmonic oscillator and yielding standard entropy $S1=1/2(1+ln(pi/beta))$ consistent with quantum statistical limit theory. At $\alpha=1$, the system acts as a fixed point of fractional information flow, where nonlocal correlations vanish, and statistical geometry reduces to a flat structure in the Fisher-Shannon information-space manifold representation. Also, for $\alpha$ less than one, the RL kernel introduces a nonlocal memory structure that redistributes probability mass away from the Gaussian core, producing algebraic tails and enlarging distribution support, reflected in entropy increase $\Delta S_\alpha$ greater than or equal to zero within the fractional entropy formulation analysis framework context. In this case, this increase in entropy corresponds to fractional delocalization, where the system transitions from an exponential confinement regime to Lévy-like heavy-tailed spreading, consistent with Fisher analysis in which the kinetic term becomes nonquadratic and produces weaker confinement at the level of spectral scaling behavior within the quantum statistical framework analysis consistency. In this formulation, Shannon and Fisher measures agree in interpretation: decreasing $\alpha$ increases uncertainty, enhances spatial dispersion, and reduces quantum state rigidity, while $\alpha$ approaching one restores maximal localization and minimal informational disorder in the system consistently within framework analysis. Moreover, entropy decomposition into geometric, scaling-anomaly, and memory-fluctuation components shows that fractional correction represents a structured modification of underlying information geometry, where logarithmic scale contribution and RL memory kernel encode distinct sources of deviation from Gaussian behavior while preserving separation of scaling and memory-induced effects in entropy space. Despite nonlocal deformation, the model preserves analytic tractability via convergent series representations and special-function forms, ensuring entropy remains well-defined across the full range $0<\alpha<=1$ within mathematical consistency conditions analysis. In this context, comparison with the fractional Fisher framework shows structural correspondence: both approaches exhibit $\alpha-$dependent weakening of confinement, identical classical limit at $\alpha=1$, and consistent interpretation via Lévy-type nonlocality and modified phase-space scaling behavior within a unified statistical description framework. Therefore, RL fractional Shannon entropy extends classical information theory and integrates consistently with fractional quantum mechanics, providing a unified description in which $\alpha$ approaching one serves as a bridge to standard Gaussian quantum statistics, while $\alpha$ less than one defines a regime characterized by nonlocal information flow, enhanced entropy production, and curved statistical geometry governed by memory-induced correlations in phase space structure.

Future investigations extend this fractional information-theoretic framework along several directions in physics contexts. A first step consists of generalizing the present RL fractional Shannon entropy to excited states of the quantum harmonic oscillator, where nodal structures and oscillatory behavior increase the sensitivity of the entropy with respect to the fractional parameter $(\alpha)$, producing information transitions between localized and delocalized regimes in phase space analysis. Also, important extension concerns higher-dimensional oscillators and anisotropic systems, where coupling between spatial directions generates direction-dependent fractional information flow and enriched entropy landscapes in quantum settings. From a more fundamental perspective, it becomes necessary to examine the thermodynamic interpretation of the fractional entropy, with emphasis on its role in defining generalized partition functions and equilibrium states in nonlocal statistical mechanics, together with its connection to entropy production in open quantum systems with memory effects. In parallel, a deeper correspondence between the RL fractional Shannon entropy and fractional Fisher information geometry may establish a unified variational principle governing both uncertainty measures through a consistent formulation on fractional probability manifolds in mathematical space. In this case, numerical studies of $(\Delta S_\alpha)$ across potentials beyond the harmonic case, including Coulomb, Morse, and relativistic oscillators, aim to clarify the universality of entropy enhancement under nonlocal deformation effects regimes. In this context, possible applications in quantum information processing are considered, where fractional memory effects may be used to model decoherence, long-range correlations, and anomalous diffusion in quantum states, enabling implementations of fractional quantum information theory in realistic physical platform contexts.

\section*{Funding}
No funding was received for this work.

\section*{Data Availability}
The datasets generated during this study can be obtained from the corresponding author upon reasonable request.

\section*{Financial Disclosure}
The authors declare no financial conflicts of interest.

\bibliographystyle{apsrev4-1}  
\bibliography{Bouzenada}

\end{document}